# Which type of citation analysis generates the most accurate taxonomy of scientific and technical knowledge?

Richard Klavans[a] and Kevin W. Boyack[b]

[a] SciTech Strategies, Inc., Wayne, PA 19087 (USA) (rklavans@mapofscience.com)

[b] SciTech Strategies, Inc., Albuquerque, NM 87122 (USA) (kboyack@mapofscience.com), Corresponding author, phone: 1-505-856-1267

## Abstract

In 1965, Derek de Solla Price foresaw the day when a citation-based taxonomy of science and technology would be delineated and correspondingly used for science policy. A taxonomy needs to be comprehensive and accurate if it is to be useful for policy making, especially now that policy makers are utilizing citation-based indicators to evaluate people, institutions and laboratories. Determining the accuracy of a taxonomy, however, remains a challenge. Previous work on the accuracy of partition solutions is sparse, and the results of those studies, while useful, have not been definitive. In this study we compare the accuracies of topic-level taxonomies based on the clustering of documents using direct citation, bibliographic coupling, and co-citation. Using a set of new gold standards – articles with at least 100 references – we find that direct citation is better at concentrating references than either bibliographic coupling or co-citation. Using the assumption that higher concentrations of references denote more accurate clusters, direct citation thus provides a more accurate representation of the taxonomy of scientific and technical knowledge than either bibliographic coupling or co-citation. We also find that discipline-level taxonomies based on journal schema are highly inaccurate compared to topic-level taxonomies, and recommend against their use.

## Introduction

In his landmark 1965 article "Networks of Scientific Papers", Derek de Solla Price (1965) made a clear distinction between research fronts and taxonomies. Price considered the research front to be the "growing tip" of the literature, comprised of about the 50 most recent papers on a topic. Others have defined the research front in different terms, but in general it is considered to represent how the current literature on a topic is being self-organized by its authors. The research front may change dramatically from year to year due to new discoveries, external events, or waning interest due to a lack of current discoveries. All of these events have the potential to change the reference frame from which a current topic is viewed. In contrast, taxonomic subjects were considered by Price to contain the rest of the papers in a topic, as well as some of those in the research front. In other words, a taxonomic subject takes into account all of the historical linkages between documents that have occurred over time, and thus reflects the history of a topic. Taxonomic subjects, and thus taxonomies (or classification systems) comprised of a number of taxonomic subjects, are inherently stable; research fronts are not. Price also noted that there were natural partitions between topics, that the research front is "divided by dropped stitches into quite small segments and strips", and that "if one would work out the nature of such strips, it might lead to a method for delineating the topography of current scientific literature (Price, 1965, p. 515)."

The past 40 years have seen a number of studies aimed at delineating the topography of the literature. Studies have been done with both small and large literature datasets, at the journal level and at the document level. Although many of these studies (including some of ours) have reported on the face validity of their results, few have attempted to measure the accuracy of their solutions in a rigorous and defensible way. Admittedly, this is difficult due to the lack of ground truth or gold standards for literature partitions. It is also true that there are many legitimate ways to classify the literature (e.g., topics/disciplines, instrumentation, basic vs. applied, etc.), and that accuracy would be defined differently for different bases. This does not, however, excuse us as a community for failing to make the addressing of accuracy a standard part of our reporting. Methods and processes for delineating topography go by multiple names including partitioning, clustering, topic detection, community detection, etc. In keeping with the spirit of Price's work, in this study we will refer to this process as "creating a taxonomy", and to sets of partitions of the literature as taxonomies.

In earlier work (Boyack & Klavans, 2010; Boyack et al., 2011) we compared the accuracy of research fronts created from over 2.15 million documents using 13 different methods. In this study, we turn our attention from research fronts to the task of accurately representing knowledge taxonomies. Document-level taxonomies created using several citation-based methods, and at several different levels of granularity, were evaluated and compared to determine which provides the most accurate representation of taxonomic knowledge. In addition, a number of journal-disciplinary taxonomies were evaluated alongside the document-based taxonomies. Journals have been used as convenient analytical constructs to define disciplines for many decades now. This is a tradition that has, perhaps, become outdated. In this paper we will show that journal-based taxonomies provide far less accurate representations of knowledge than do document-level taxonomies. This result suggests that journal-based taxonomies are a poor basis for research evaluation and for the measurement of interdisciplinarity, and that they should be replaced by topic-level taxonomies.

As was done in our previous study, we introduce a new metric for assessing accuracy. We propose that papers with at least 100 references can be considered as gold standards for taxonomic subjects, and that the concentration of their references can be used to evaluate and compare different methods for creating taxonomies. Writing a paper with 100 references is a daunting task – only 3.14% of all articles and reviews published in 2010 reached this level of comprehensiveness. Nearly half (48.1%) of these papers are identified as review papers by Scopus, and those that are not have the same general characteristics as the review papers. Papers with 100 references are more likely to be written by authors that have published continuously for seven years than papers with fewer than 100 references. Further justification for using papers with 100 references as gold standards will be given below, along with a detailed characterization of this corpus. The following sections will provide a historical background for the present work; detail the various processes, models, and standards to be used; provide a detailed analysis of results, and explore associated implications.

## Background

### *Early citationists*

Citation analysis, and the use of citation-based methods – direct citation, co-citation, and bibliographic coupling – to understand the structure of science, have a rich history. The initial goal of this research stream was to detect emerging research communities as they are emerging in order to redirect resources towards these opportunities. Technical barriers (e.g., lack of comprehensive electronic data, and lack of computing resources) kept early researchers from reaching this goal. It is only recently that creation of comprehensive taxonomies of the literature has become possible. The intervening 50-year period represents two generations of academics who may not be aware of the historical context of this research stream. Accordingly, in this section we revisit these issues and place document-level citation analysis in its historical context.

The most appropriate place to start is with Garfield's (1955) influential article on "Citation Indexes for Science". He was particularly concerned that, when analyzing the scientific literature, subject indexes could not pick up an emerging area of science as it was emerging. Citation analysis was introduced as a more effective way to index the literature. In his example, he starts with Selye's seminal article on general adaptation syndrome[i] and tracks its impact using the 23 articles from the same journal over the subsequent five years that cite this paper. Garfield noted that, even within this same journal, the citing articles were not associated with the same subject code. When Selye's paper was published in 1946, and for the next few years thereafter, a subject heading for general adaptation syndrome (GAS) didn't exist. In addition, early publications on GAS were dispersed among multiple journals. This example provided evidence that citations might be useful in tracking an emerging concept. Garfield was also highlighting the problem with using subject headings to detect a breakthrough in real time.

Nearly a decade later, Garfield et al. (1964) proposed that direct citation analysis could be used to trace the historical development of a scientific breakthrough. Isaac Asimov's 1964 book on the discovery of DNA code[ii] was used to identify the 65 most important papers associated with this discovery. The bibliographies of these papers were collected and analyzed to determine 'who cites whom'. 65% of these papers were connected by direct citation links. Additional links could be discovered using interim papers. More importantly, there was strong agreement between Asimov's interpretation of historical events and a temporal layout of the citation network. This example provided evidence that direct citation analysis could be used to create a single coherent category of knowledge.

At approximately the same time, Kessler (1963) proposed the use of bibliographic coupling to reveal the structure within physics. Kessler was not looking at historical links. He was taking a snapshot in time with the intent of showing the research front. The difference between direct citation analysis and bibliographic coupling was highlighted by Price (1965), who considered direct citation analysis as taxonomic – a systematic structure that "treats each published item as it were truly part of the eternal record of human knowledge". In contrast, bibliographic coupling reveals what scientists are currently working on. Research fronts will change from year to year as researchers make discoveries and shift their attention to different research problems.

Kuhn's (1962) book on the structure of scientific revolutions was also published at around this time. The response to this book was explosive – there was very broad interest in scientific revolutions in society at that time. It is important to note that Kuhn and Price knew of each other's work and were known to be competitors. But despite the intellectual divide between Kuhn and Price, there is a very telling quote in the 2nd edition of Kuhn's book (1970) where he elaborates on the concept of a research community:

> "How, to take a contemporary example, would one have isolated the phage group prior to its public acclaim? For this purpose one must have recourse to attendance at special conferences, to the distribution of draft manuscripts or galley proofs prior to publication, and above all to formal and informal communication networks including those discovered in correspondence and the *linkages among citations [6]*. I take it that the job can and will be done." (Kuhn, 1970, p. 178, emphasis added)

Kuhn was raising the same question that Garfield had raised earlier – how could one have detected this scientific revolution as it was emerging? His commentary specifically mentioned "linkages among citations" and his footnote [6] was to three articles: Garfield et al. (1964), Kessler (1965) and Price (1965). For the period between 1955 and 1972, we think it is fair to conclude that one of the primary motivations for citation analyses was to reveal scientific breakthroughs as they were occurring. This was a major interest voiced by citationists (Price, Garfield, and Kessler) and the most influential science historian of the 20th century (Kuhn). Furthermore, Kuhn assumed that "the job can and will be done". Forty-five years have passed since this optimistic view was taken, and it has taken all of that time to get to a point where we, as a community, can say that we're getting close. The technical barriers (i.e., lack of comprehensive electronic data, lack of computing resources, and lack of gold standards) that have kept us from reaching this goal are just now being overcome.

### *Constrained progress*

Have you ever wondered why direct citation wasn't the first approach used to create a knowledge taxonomy for the entire ISI database? Eugene Garfield was the Founder and President of ISI, the only source of citation data covering all of science until 2004. He was also a strong advocate of direct citation analysis as evidenced by his interest in historiography and his many case studies, and likely had the influence to make direct citation a standard in the field. Nevertheless, the first approach to creating a taxonomy of all of science using citation data was the co-citation model introduced by Small & Griffith (1974). In fact, the first taxonomy covering all of science using direct citation was not created until nearly 40 years later, by Waltman & van Eck (2012). Why did it take so long?

A look at the history of studies reporting on document-based taxonomies of all of science (see Table 1) reveals a very interesting trend. Figure 1 shows that the relationship between the log of the number of papers and publication year for these studies is surprisingly linear, suggesting a technical barrier of sorts to the creation of large taxonomies. We posit that this technical barrier, which changed over time, was related to computational capabilities, and more specifically to the computer memory needed to cluster documents sets. One might argue that supercomputers were available during the 1980s and 1990s that could have clustered millions of documents. Although this is certainly true, these computers were extremely costly, and were not available to the researchers creating taxonomies of the scientific literature. Rather, these researchers were limited to use of desktop computers and small mainframe computers for their calculations. We further posit that this limitation in computing resources played a role in the choice of methodology used to create taxonomies of science, and will explain that position in the following discussion of each of the studies listed in Table 1.

**Table 1. Characteristics of document-level taxonomies of all of science.**

| Publication | # Nodes | Method | Algorithm |
|---|---|---|---|
| Small & Griffith (1974) | 1,832 | CC | Single-link |
| Small et al. (1985) | 43,931 | CC | Single-link |
| Franklin & Johnson (1988, ISI) | 72,077 | CC | Single-link |
| Franklin & Johnson (1988, CRP) | 195,036 | CC | Single-link |
| Small (1999) | 164,612 | CC | Single-link |
| Klavans & Boyack (2006) | 731,289 | BC,CC | VxOrd |
| Boyack (2009) | 997,775 | BC | VxOrd |
| Klavans & Boyack (2010) | 2,080,000 | CC | DrL/OpenOrd |
| Waltman & van Eck (2012) | 10,200,000 | DC | Smart local moving |
| Boyack & Klavans (2014c) | 43,431,588 | DC | Smart local moving |

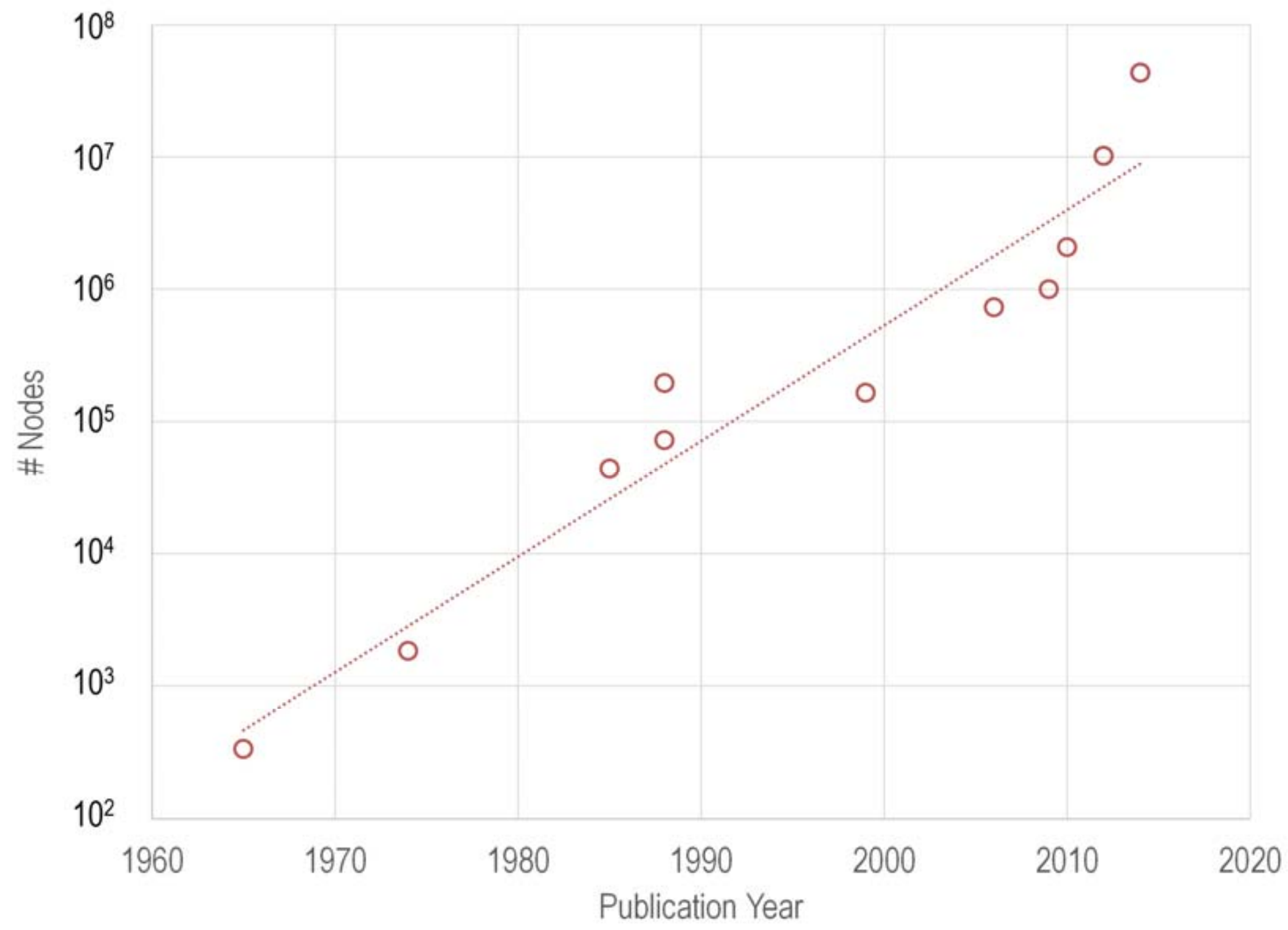

**Figure 1. Progress in the creation of global taxonomies and research fronts.**

The first point in Figure 1 represents Kessler's (1965) bibliographic coupling analysis of structure in physics using 334 documents. Although this analysis did not incorporate papers from all of science, it was nonetheless the largest literature 'clustering' study of its time, and is a suitable departure point for discussion of taxonomy sizes. All subsequent taxonomies of specific fields, disciplines or topics are well below the trend line (or barrier) shown in Figure 1.

The next point in Figure 1 represents Small & Griffith (1974), who created the first cluster solution covering all of science. One might wonder, however, why co-citation analysis was chosen for this work over direct citation and bibliographic coupling. In hindsight, there appear to be several things that may have contributed to this decision. First, a single year of data was far too large to cluster using the technology available at that time; the 1972 SCI data contained 93,800 source documents and 867,600 references. A direct citation calculation using all links would have required clustering of not only the 93,800 source papers from 1972, but also another roughly 300,000 cited papers (assuming a standard ratio of single year cites to papers of 3:1). Clustering nearly 400,000 papers was inconceivable at that time; thus, it was clear that some sort of drastic thresholding of the data would be necessary.

Second, we need to remember that the prevailing motive for creating a taxonomy of science at that time was to identify emerging topics, and that it was assumed by the early citationists that direct citation would work for that task. Emerging topics make up only a small fraction of the literature at any one time. Thus, the question became how best to threshold the data so that one could create a taxonomy of the emerging topics across all of science. Thresholding from a direct citation perspective is extremely difficult; one doesn't know *a priori* which papers from the source year are the most important, and even if one did, the direct citation network is so sparse that drastic thresholding is likely to fully disconnect the important papers. Using this logic, direct citation was not a viable choice at the time for creating a taxonomy of scientific knowledge. This left co-citation and bibliographic coupling as possible choices.

For Henry Small, the choice between these two methods was a simple one. In his introduction of the co-citation method, Small (1973) had investigated this question by generating the links between each pair among 10 physics papers using direct citation, co-citation and bibliographic coupling. An analysis of the link strengths suggested that co-citation analysis was very similar to direct citation, while bibliographic coupling was not. Further analysis of Small's 1973 data shows that the correlation between co-citation and direct citation (r=0.50) was indeed higher than that between bibliographic coupling and direct citation (r=0.16). Given its similarity to direct citation, and given the desire to identify emerging topics, one could expect that co-citation would create a reasonable taxonomy of research. Furthermore, if one assumes that emerging topics are the same as the most important topics, then the thresholding task becomes tractable. One simply chooses the most highly cited papers and creates the taxonomy using that set. Accordingly, Small & Griffith (1974) identified a set of 1,832 highly cited papers, calculated 20,414 co-citation links between those papers, and used single-link clustering to create a set of 115 clusters containing 1,310 of the highly cited papers. The remaining papers were not linked to a cluster. Once these highly co-cited references were clustered, citing papers could easily be assigned to those clusters, thus identifying the "research fronts" or the current activity associated with emerging or important topics across science. In essence, Small & Griffith reduced the magnitude of the computational problem from one requiring hundreds of thousands of nodes and links to a much smaller set, and in doing so, set co-citation analysis as a standard that would be used for many years to come.

The next four examples in Table 1 (Franklin & Johnston, 1988; Small, 1999; Small et al., 1985) and Figure 1 represent significant improvements in computational capabilities. Here we focus on the two from 1988. Both of these examples are taxonomies of the ISI databases created using co-citation analysis, and the two resulting models underwent a thorough review by outside experts (Franklin & Johnston, 1988). In 1983, Henry Small had increased the number of highly cited references from 1,832 to 72,077 (of which 50,994 ended up being clustered). Len Simon of the Center for Research Planning (CRP), also using the ISI database and co-citation analysis, increased the number further to 195,036 highly-cited references (of which 128,238 ended up being clustered). The reasoning for these differences is straightforward – Small only wanted to identify the hottest science, while Simon was interested in hot, warm and colder topics. Over the next 10 years, Simon continued to increase the number of reference papers so that one could identify hot, warm and cold topics. Small maintained a high threshold in order to focus solely on potential scientific breakthroughs.

These two co-citation models were not without problems. One problem was the reliance on single link clustering. This was the clustering algorithm of choice because it was relatively simple to use, and using common computers of the day could handle document sets of around a hundred thousand papers. However, it is well known that single-link clustering creates over-aggregated clusters due to chaining effects. Small and Simon were both aware of this problem – Small overcame the problem by imposing a maximum cluster size, while Simon avoided using the largest clusters until he could develop a way to break them up into meaningful groups, which took 5 years of work.

Advances in clustering algorithms came from unexpected directions. The first modularity-based algorithms capable of clustering hundreds of thousands of documents became available in the mid-2000s (Newman, 2006; Newman & Girvan, 2004). In addition, the VxOrd algorithm from Sandia National Laboratories (Davidson, Wylie, & Boyack, 2001), which had originally handled only tens of thousands of nodes, was revamped (and renamed to DrL, and then to OpenOrd) to where it was capable of clustering millions of nodes (Martin, Brown, Klavans, & Boyack, 2011). OpenOrd is a force-directed placement layout algorithm that employs edge cutting. When the cutting parameter is set to a maximum, the outcome is a layout with well-articulated groups of nodes. Use of single-link clustering on OpenOrd output creates clusters without chaining. This algorithm and procedure were used to create the next three taxonomies from Table 1 and Figure 1 (Boyack, 2009; Klavans & Boyack, 2006, 2010). For the largest of these taxonomies, 2.08 million references were clustered using co-citation analysis. The number of edges used in these co-citation models was typically about a factor of 10 larger than the number of nodes, thus around 20 million edges in this case. From the standpoint of a technical barrier, 2 million nodes and 20 million edges are significant numbers in that this calculation could be done using OpenOrd on a standard desktop computer (8GB RAM) within memory. Larger calculations (e.g. 3 million nodes, 30 million edges) would overrun memory and require disk swap space, which made the calculation so slow as to be useless.

The final two studies represent a sea change in the creation of taxonomies of the S&T literature, and indeed, the final two data points in Figure 1 are well above the historical trend line. Waltman & van Eck (2012) from CWTS introduced a new modularity-based clustering algorithm and showed its utility by classifying 10 million papers from a 10 year segment of the Web of Science database using 97 million direct citation links. After testing, we subsequently used the same algorithm on a much larger set of 43.4 million documents and 510 million direct citation links from 16 years of Scopus data. This most recent taxonomy includes 22 million non-source-indexed documents that were cited at least twice, and represents the largest taxonomy of the S&T literature published to date (Boyack & Klavans, 2014c). Note, however, that this taxonomy was not created on a desktop computer or small server. Rather, the calculations were performed on the Amazon Compute Cloud at a cost of only a few hundred dollars. Thus, it is not only advances in algorithms, but also the recent availability of low cost, high-performance computing that enables the capability we have today to create comprehensive taxonomies of tens of millions of documents.

Over 40 years ago a lack of computing capacity constrained the ability of our citationist pioneers to create a comprehensive taxonomy covering all of science. This ultimately did not deter them from trying to solve the problem, but led to choices (thresholding and co-citation) that allowed progress to be made. Since that time, the technical barrier suggested by Figure 1 has been consistently eroded, and we are now at a point where data, rather than algorithms and computing capability, may be the limiting factor. Nevertheless, we envision the day will shortly come when a taxonomy will be created from well over 100 million documents (including papers, patents, grants, web pages, etc.)

Now that this barrier has been overcome, we still need to answer the original question – can these taxonomies identify emerging areas as they are emerging? One recent study showed that a combination of two models (a taxonomic model based on direct citation, and a co-citation model to detect research fronts) nominated emerging areas as they were emerging with negligible false positives (Small, Boyack, & Klavans, 2014). However, we still don’t know which citation-based strategy creates the most accurate taxonomy of knowledge. This issue is discussed in the following section.

### *Accuracy and gold standards*

In most fields of science, accuracy is of paramount concern. Admittedly, some fields lend themselves more to accuracy than others. This is particularly true for those fields where physical properties can be measured, those for which gold standards exist, and those where a great deal of research is replicated. Unfortunately, none of these conditions are extant when it comes to the delineation of topics, or the creation of taxonomies of the scientific literature; there is nothing physical to measure, there are no gold standards, and relatively little research is replicated. In addition, one must also consider that a taxonomy created by any particular process may, in fact, contain artifacts that are dependent on that process itself (Leydesdorff, 1987). For example, a recent analysis of community detection algorithms found that the communities produced by these algorithms are in many cases more related to network topology than to the underlying real-world feature space (Hric, Darst, & Fortunato, 2014).

Given this state of affairs, most studies clustering the literature have effectively ‘thrown up their hands’ and have relied on face validity rather than some objective measure of accuracy. Objectivity is also difficult to achieve. The majority of studies that have provided some measure of accuracy (a review of these studies can be found in Boyack & Klavans (2010)) have either compared among similarity measures, or have compared results to a well-known basis, such as the WoS subject categories. Although admirable and useful to an extent, none of these comparisons are truly objective.

Perhaps the most objective study on accuracy of different literature taxonomies is the recent work by Boyack and colleagues (Boyack & Klavans, 2010; Boyack et al., 2011), in which the relative accuracies of research fronts created from over 2.15 million documents using 13 different methods – three citation-based methods, nine text-based methods, and one hybrid method – were compared. The key to that study was the use of a novel accuracy metric, the concentration of grant-to-article linkages from PubMed. The working assumption behind this metric was that articles acknowledging a single grant would be topically similar, should thus be concentrated in the cluster solution, and that cluster accuracy should therefore correlate positively with grant concentration. Comparisons were done at the topic level (tens of documents per cluster per year) rather than the discipline level (thousands of documents per cluster per year) (Boyack, Klavans, Small, & Ungar, 2014). The results were considered robust given the large size of the corpus and the objectivity – the grant-to-article data did not directly contribute to the different cluster solutions in any way – by which accuracy was determined.

In this study we build upon our earlier work. However, rather than looking at short-term research fronts, we are interested in the accuracy of much longer-term taxonomies, and in determining which citation-based process creates the most accurate taxonomy. We acknowledge that absolute accuracy is a fleeting objective, and given that topic detection is not completely free from artifact, is not possible. Nevertheless, an accurate structure in terms of how that structure is perceived by researchers, administrators, and funders, is necessary to enable the best possible decisions by those actors. To that end, we have compared nine different global document-based taxonomies and seven different global journal-based taxonomies. The following sections characterize each of those taxonomies, define a new metric for accuracy, provide a detailed analysis of the results, and explore associated implications.

## Methods and Models

### *Citation-based taxonomies*

Our previous comparison of the three citation-based methods (Boyack & Klavans, 2010) came to the conclusion that bibliographic coupling (BC) was the most accurate, followed by co-citation (CC). Direct citation (DC) was a distant third among the three. However, that study was based on a short time window of 5 years, which puts DC at a distinct disadvantage in comparison with CC and BC. After learning about the modularity-based algorithm of Waltman & van Eck (2012), we replicated their study, and found that the accuracy of their 10-year DC solution was comparable to that of our previous CC and BC solutions.[iii] This confirmed to us that the short time window in our previous study was indeed the reason that DC did not appear to be very accurate.

In this study, we have generated taxonomies based on all three citation-based processes (DC, CC, and BC) for comparison. The mechanics of and differences between these three citation-based processes have been amply explained elsewhere (cf. Boyack & Klavans, 2010), and will not be reviewed in detail here.

The CWTS modularity-based method and optimization algorithm, although it is commonly associated with direct citation by the science mapping community because of the way it has been used to date, can be used with any of the three citation-based processes, or even with text-based processes, because it uses link strengths between pairs of nodes as input. All of the taxonomies used in this study were thus created using the CWTS method. Details associated with the nine taxonomies and their creation are given in Table 2. The CWTS method is capable of creating hierarchical solutions within a single calculation run. Although nine solutions are listed, only three calculations were done – one each for DC, CC, and BC. The DC and BC calculations were run at four levels each – resolutions and minimum cluster sizes were set to produce solutions with approximately $10^2$, $10^3$, $10^4$, and $10^5$ clusters.

**Table 2. Characteristics of nine document-based taxonomies of Scopus data. Superscripts in model names reflect the order of magnitude of the number of clusters (e.g. $DC^5$ has roughly $10^5$ clusters).**

| Type | Source Years | Taxon | Res | Min Size | #Clust | #Doc | #Base | #Doc ≤ 2009 | Rel Cov |
|---|---|---|---|---|---|---|---|---|---|
| DC | 1996-2012 | $DC^2$ | 3e-8 | 50,000 | 114 | 46,705,047 | | 40,186,275 | 96.5% |
| | | $DC^3$ | 3e-7 | 5,000 | 849 | 46,929,303 | | 40,384,510 | 97.0% |
| | | $DC^4$ | 3e-6 | 500 | 10,059 | 47,323,189 | | 40,722,075 | 97.8% |
| | | $DC^5$ | 3e-5 | 50 | 91,726 | 48,533,301 | | 41,639,235 | 100.0% |
| CC | 2011-2013 | $CC^5$ | 1.1e-5 | 30 | 92,571 | 18,095,283 | | 14,905,085 | 35.8% |
| BC | 2010 | $BC^2$ | .0064 | 2,000 | 109 | 2,037,829 | 7,974,601 | 7,785,893 | 18.7% |
| | | $BC^3$ | .0064 | 200 | 1,070 | 2,037,829 | 7,419,150 | 7,245,513 | 17.4% |
| | | $BC^4$ | .0024 | 20 | 10,548 | 2,037,829 | 6,859,954 | 6,698,097 | 16.1% |
| | | $BC^5$ | .0014 | 2 | 101,337 | 2,044,538 | 5,739,526 | 5,599,379 | 13.4% |

***Direct citation:*** The DC calculation was based on the entire set of Scopus indexed source documents from 1996-2012, and also included non-indexed documents cited at least twice. All source documents were used because it has become clear that DC performs best when using long time windows, and our intention is to determine the accuracy of historical taxonomies. The detailed methodology used to generate the document set and citing-cited pairs list is very similar to that used for the recent model of Boyack & Klavans (2014c). Input for the DC calculation was comprised of 48.5 million documents (24.6 million source, 23.9 million non-source) and over 582 million citing-cited pairs. Direct citation produced the highest coverage of all citation-based methods, leaving very little of the important historical literature uncovered (or obliterated) when cited non-source documents are included. The updated version of the CWTS method, which now uses a smart local moving (SLM) algorithm for optimization (Waltman & van Eck, 2013) – was used to create this set of taxonomies. Table 2 shows the numbers of clusters and documents in each of the DC taxonomies. The number of documents is reduced as one moves up the hierarchical solution due to clusters dropping out (not being linked to other clusters) upon aggregation.

***Bibliographic coupling:*** The BC calculation was based on a single-year window (2010) of Scopus source documents. A single year was chosen in this case because we wanted a taxonomy that reflected which cited documents were viewed to be "important" in a single year, to see if the "current importance" angle might prove to be a more accurate view than the full history. In a sense, bibliographic coupling obliterates history when it is based on a short time window. 2010 was chosen because we have done a variety of other (unpublished) calculations with the 2010 file year, and were thus more familiar with this file year than others. Bibliographic coupling was done using the 2,044,538 documents from 2010 with references using the process detailed in Boyack & Klavans (2010). Coupling using all references resulted in a set of 818 million edges. The number of edges was reduced to 197 million by excluding all references from the bibliographic coupling coefficients that had been cited more than 100 times in 2010. Once the 2010 papers were clustered, cited references were assigned to their dominant cluster (see the "#Base" column in Table 2) for cases where at least 40% of the citations to the cited reference were from 2010 papers in a single cluster. Cited references with ambiguous assignments (ties), a dominant cluster containing less than 40% of the citations, or cited only once were not assigned. Note that in comparison with $DC^5$, the bibliographic coupling taxonomies omit tens of millions of documents. This can be viewed as a natural process of obliteration; only documents that are relevant to current science are included in these taxonomies.

***Co-citation:*** The CC calculation was based on a three-year window (2011-2013) of Scopus source documents. In the past we had used single-year windows due to computational constraints. Given the ability of the SLM algorithm (and use of the Amazon Compute Cloud) to handle hundreds of millions of edges, we shifted to a three-year window for this calculation. It was also hoped that this would create a more stable co-citation solution than is obtained from a single-year window. Nearly 18.1 million documents were cited at least twice during the three-year window, and co-citation similarities were calculated for each pair of these co-cited documents using the methodology in Boyack & Klavans (2014b). Rather than using the full set of 3.3 billion edges, we calculated the top-n (range 5-15) edges per paper, and used this filtered set of 111 million edges as input to the SLM algorithm. The resolution and minimum cluster size values were set to create approximately $10^5$ clusters so that the resulting taxonomy could be compared to the $DC^5$ and $BC^5$ taxonomies. Our initial analyses showed the co-citation clustering at the $10^5$ level was so much less accurate than the corresponding DC and BC solutions that we decided not to create more aggregated CC solutions.

### *Journal-disciplinary taxonomies*

In this study, we evaluate several journal-based taxonomies alongside the document-level taxonomies that we have created. This is an important comparison to make given that journal-based taxonomies are pervasive and continue to be used worldwide for a number of evaluative purposes. Seven journal classification schemes have been located and linked to Scopus data through journal names, namely:

- Elsevier's All Science Journal Classification (ASJC), used in Scopus[iv]
- UCSD journal classification (Börner et al., 2012)[v]
- Science-Metrix (SM) journal classification (Archambault, Caruso, & Beauchesne, 2011)[vi]
- Australian Research Council (ARC) journal classification[vii]
- KU Leuven ECOOM journal classification (Glänzel & Schubert, 2003)[viii]
- Web of Science subject categories[ix]
- U.S. National Science Foundation (NSF) journal classification, used in the biannual Science & Engineering Indicators reports[x]

Not all journals in all schemes are available in Scopus. The number of source titles (journals and/or conference proceedings) from each scheme that were matched to Scopus titles is shown in Table 3, along with numbers of journal categories, and the number and fraction (relative coverage) of papers from the $DC^5$ document-level taxonomy that were located in each journal-based taxonomy. The UCSD, SM, and NSF schemes assign journals to single categories, while the ASJC, ARC, ECOOM and WOS schemes assign some journals to multiple categories. Note that for ASJC, both top-level and bottom-level categories were used. However, top-level categories were only used for those journals (and thus papers) that were not assigned to a bottom-level category.

In addition to the seven journal classification systems, we also consider the situation where each journal (or conference) is considered as its own category. Although use of journals as knowledge categories is not yet a commonplace occurrence, this method is nonetheless being used at a sufficient frequency that it should be considered. For instance, Uzzi et al. (2013) used journals to represent knowledge categories to show how atypical knowledge relationships (pairs of co-cited journals) can be used as an indicator of innovativeness, while Leydesdorff et al. (2013) use individual journals in an overlay map of science. In this study, we use Scopus journal-IDs (JID) to represent the taxonomy of individual journals.

**Table 3. Characteristics of seven journal-based partitions of Scopus data.**

| Name | #Jnl/Src | #Cat | #Doc thru 2009 | Rel Cov |
|---|---:|---:|---:|---:|
| ASJC | 37,635 | 332 | 25,048,976 | 60.0% |
| UCSD | 21,582 | 547 | 21,753,045 | 52.1% |
| Science-Metrix | 14,866 | 176 | 20,604,866 | 49.4% |
| ARC | 15,615 | 173 | 18,605,882 | 44.6% |
| ECOOM | 10,543 | 68 | 17,967,621 | 43.0% |
| WOS | 12,420 | 251 | 17,822,758 | 42.8% |
| NSF | 7,918 | 138 | 16,902,805 | 40.5% |
| JID (journals) | 37,635 | 37,635 | 25,048,976 | 60.0% |

ASJC and JID have the highest coverage of the seven journal classification schemes, which is not surprising given that they represent the system used in Scopus. Their relative coverages are only 60% because journal IDs are not assigned to the majority of the non-source items that are included in the DC[5] taxonomy. UCSD has the second highest coverage, which is logical in that it was developed using lists of journals from both Scopus and the Web of Science. However, approximately 40 multidisciplinary journals (e.g., Science, Nature, PNAS, Cell, etc.) are not included in this scheme because they were not singly assigned to UCSD categories (Börner et al., 2012). The NSF scheme has the lowest coverage, which is not surprising because this list was based on the original Science Citation Index and held constant for many years to facilitate year-to-year comparisons of output in the biannual S&E Indicators reports. Thus, newer prominent journals (including the PLOS journals) were not included in the 2005 version of this scheme that was available to us.

### *Proposed gold standards and accuracy*

The above sections describe the 17 taxonomies, or models of science, to be compared. The question now is how best to compare them to determine which provides the most accurate representation of the taxonomy of scientific and technical knowledge. Recall that our intent is to measure the accuracy of historical silos (Price's taxonomic subjects) of knowledge rather than current snapshots (research fronts). The accuracy metrics we have used in the past (Boyack & Klavans, 2010) seem less well suited to the task than they did for comparing research fronts. Textual coherence, while it is a powerful measure of the tightness of language used in a cluster, seems less valid in a historical setting where topics can shift their focus and topic breadth can change over time. Grant-to-article links remain a good choice for biomedicine, but are less well suited for all of science because very few grant-to-article linkage data exist outside biomedicine.

As mentioned above, there are no agreed-upon gold standards (defined as examples of ground truth) for literature partitions. For this study, we propose that papers with at least 100 references can be considered as gold standards for taxonomic subjects, and that the concentration of their references can be used to evaluate and compare different methods for creating taxonomies. There is historical, as well as current, justification for this view. From the historical perspective, Price (1965, p. 515) opined about the need for a review paper after every 30-40 papers to summarize "those earlier papers that have been lost from sight behind the research front." Implicit in this argument is the assumption that a review paper will be focused on a specific topic. Guides to writing literature reviews often give similar advice. For example, one suggestion in a recently published list of rules for writing a literature review (Pautasso, 2013) is to "keep the review focused, but make it of broad interest."

Studies have acknowledged the special roles that review articles play in the fabric of science. Woodward (1977) identified two interlocking roles for review articles – a contemporary function that informs researchers about current research, and a second function that weaves that development into an historical whole. In this sense, Woodward's view mirrors that of Price's research fronts and taxonomic subjects, and suggests that review articles combine the two. Reviews are also known to have more references and to be more highly cited, on average, than articles reporting on original research (Aksnes, 2003).

Ultimately, however, the idea of using papers with extremely long bibliographies as gold standards for partitions came to us from current research. We recently conducted a survey of the most elite authors in biomedical research in which we asked them to rank motives (innovativeness and synthesis were two of these motives) for their top 10 most highly cited recent papers. We expected that there would be a tradeoff between innovativeness and synthesis. What we found, however, was that these elite authors considered their synthesis papers to be just as important as their papers describing innovative original research (Ioannidis, Boyack, Small, Sorensen, & Klavans, 2014). The fact that elite authors view synthesis as some of their most important work suggests that care is taken in the construction of their review articles (and other articles with long reference lists), and that this synthesis is purposeful and topic-focused.

This experience also suggests to us that when considering the structure of science we should focus more on synthesis than review. How can one differentiate between a document that is classified as a review article, and one that truly synthesizes current literature on a topic with its historical past? Given that the classification of a document as an article or a review is not standardized – it varies by journal and database, for example – we prefer to define synthesis papers as those with large numbers of references, regardless of their database designation as an article or review.

We therefore analyzed the 2010 publications in the Scopus database with these issues in mind. Analysis was restricted to articles and reviews, and publications were binned using the number of references (1-9, 10-19, 20-29…), with increased bin sizes over 100 references. The data in Table 4 suggest three reasons why documents with at least 100 references are acting as synthesis articles and, as such, could be considered as gold standards for topical partitions. First, documents with 100+ references comprise 3.14% of the sample, or one of every 32 documents being published. This is roughly the rate of summary articles anticipated by Price (1965). Second, almost half (48.1%) of these documents with 100+ references are coded as review papers (the expected value is only 7.8%). Third, the percentage of influential core authors – those who had published continuously for at least 7 years (Ioannidis, Boyack, & Klavans, 2014; Price & Gürsey, 1975) – is highest for papers with 100+ references, while the actual number of authors per paper is very low. Author collaboration peaks in papers with 30-39 references, and then goes down to a level where the number of authors is about the same as a paper with less than 15 references.

**Table 4. Characteristics of articles and reviews from 2010. Number of citations (# Cit) and percent uncited papers (% Nocit) are as of the end of 2012.**

| # Ref | # Doc | # Ref | # Cited | # Authors | % Core Authors | % Uncited | % Review |
|---|---|---|---|---|---|---|---|
| 1-9 | 140587 | 6.2 | 1.0 | 3.2 | 25.3% | 64.2% | 4.3% |
| 10-19 | 329046 | 14.6 | 2.4 | 4.1 | 31.8% | 42.5% | 3.6% |
| 20-29 | 324540 | 24.4 | 4.4 | 4.6 | 36.4% | 26.2% | 4.5% |
| 30-39 | 255193 | 34.1 | 6.1 | 4.9 | 38.5% | 18.5% | 5.2% |
| 40-49 | 167729 | 44.1 | 6.9 | 4.8 | 38.9% | 14.7% | 7.1% |
| 50-59 | 101168 | 54.0 | 7.6 | 4.6 | 38.9% | 13.5% | 10.4% |
| 60-69 | 58368 | 64.1 | 7.8 | 4.4 | 38.5% | 13.4% | 14.3% |
| 70-79 | 35090 | 74.1 | 8.4 | 4.1 | 39.4% | 13.6% | 19.5% |
| 80-89 | 22133 | 84.1 | 8.9 | 3.8 | 39.4% | 14.0% | 25.8% |
| 90-99 | 14622 | 94.2 | 9.7 | 3.7 | 40.5% | 13.8% | 33.3% |
| 100-199 | 39843 | 129.9 | 13.6 | 3.4 | 43.4% | 13.7% | 47.6% |
| 200+ | 7107 | 302.5 | 22.6 | 3.3 | 46.3% | 16.0% | 51.1% |
| Total | 1495426 | 34.8 | 5.1 | 4.4 | 35.6% | 28.4% | 7.8% |

There is also a different relationship between citation rates, number of authors, percentage of elite authors and number of references for documents with more than and less than 100 references. The correlation matrix for documents with less than 100 references (Table 5) shows that future citation rates are highly correlated with the number of references (log transforms are used to reduce the skewness in the data). A Pearson correlation of .4034 is extremely high given the number of observations (over 1.4 million). However, for papers with at least 100 references (Table 6), the impact on the number of future citations drops dramatically (a Pearson correlation of only .0793 vs. .4034). At this threshold, the involvement of elite authors is a better predictor of future citation rates (the Pearson correlation increased from .2677 to .3887).

**Table 5. Correlation matrix for articles and reviews from 2010 with fewer than 100 references (n>1.4 million).**

| | log(# Cit+1) | log(# Ref) | log(# Auth) | % Core |
|---|---|---|---|---|
| log(# Cite+1) | 1.0000 | | | |
| log(# Ref) | 0.4034 | 1.0000 | | |
| log(# Auth) | 0.3121 | 0.1162 | 1.0000 | |
| % Core | 0.2677 | 0.1247 | 0.1687 | 1.0000 |

**Table 6. Correlation matrix for articles and reviews from 2010 with at least 100 references (n=45,457).**

| | log(# Cit+1) | log(# Ref) | log(# Auth) | % Core |
|---|---|---|---|---|
| log(# Cite+1) | 1.0000 | | | |
| log(# Ref) | 0.0793 | 1.0000 | | |
| log(# Auth) | 0.3329 | -0.0790 | 1.0000 | |
| % Core | 0.3887 | -0.0095 | 0.1809 | 1.0000 |

Using the data and logic detailed above, we have decided to use the 37,207 articles and reviews published in 2010 with at least 100 references, and for which at least 80% of those references are available in the DC5 taxonomy, as our gold standards for literature partitions. These represent roughly one synthesis paper for each 35 articles, as suggested by Price. Half are independently coded as review articles, and most have core (influential) author involvement. It is reasonable to assume that core authors know and understand the literature on a topic, and that the articles with large numbers of references written by these core authors can thus be considered as expert-based partitions of the literature. In addition, the large numbers of references in each paper provide a substantial definition of which papers belong in a particular partition, much more so than papers with only 30-40 references, thus enabling precise evaluation of the accuracy of the different taxonomies detailed above.

The 37,207 gold standard articles and their references are widely distributed throughout the sciences. These articles are located in 15,725 different $DC^5$ clusters, and their 5,334,016 references are to 3,248,243 unique reference papers (2,781,929 source, 466,314 non-source) that are located in 79,354 of the 91,726 $DC^5$ clusters. 71.2% of the reference papers in the union set are referenced by only one gold standard article, and an additional 22.1% of the reference papers are referenced by either two or three gold standard articles. Thus, the inherent overlap in the reference paper set is small, and the results to be shown below reflect contributions from all of science rather than just one field.

The relative accuracies of the 17 taxonomies detailed above are compared using a standard Herfindahl index. The Herfindahl index is an appropriate measure for this comparison in that it can be used to measure concentration of references in a taxonomy, and has a natural range from 0 to 1. Assuming that there are P gold standard papers ($p = 1 \ldots P$), the Herfindahl value for each paper $p$ in each taxonomy $i$ can be calculated as:

$$H_i^p = \sum (s_{ij}^p)^2 \ ,$$

where the standard share value for paper $p$ in cluster $j$ of taxonomy $i$ is calculated as $s_{ij}^p = n_{ij}^p \big/ N^p$ , $n_{ij}^p$ is the number of references from paper $p$ in cluster $j$ of taxonomy $i$, and $N^p$ is the total number of references in paper $p$ that are available in the $DC^5$ taxonomy. This particular taxonomy is used as the baseline because it has the highest coverage (95.7% of the references in the 37,207 gold standard articles are located in this taxonomy), and all other taxonomies are subsets of $DC^5$. Missing values – references that are not available in a particular taxonomy – are assumed to belong to their own cluster, and are thus accounted for in all taxonomies. For those journal-based taxonomies where some journals (and thus some references) are assigned to more than one cluster, values of $n_{ij}^p$ are based on fractionalizing references among the participating clusters. The overall index for a taxonomy is then calculated as:

$$H_i = \frac{1}{P} \sum_p H_i^p \ .$$

## Results and discussion

### *Taxonomy-level results*

Figure 2 shows the Herfindahl values, $H_i$, for each of the 17 taxonomies. The first thing we observe is that the DC and BC taxonomies (which are hierarchical) both show strong relationships between concentration and the number of clusters, suggesting that taxonomies with different numbers of clusters should not be compared directly. Rather, comparisons should be made between taxonomies with roughly the same numbers of clusters – i.e. in vertical silos. DC taxonomies are more concentrated than BC or CC taxonomies at all levels, and thus should provide the most accurate document-level representation of taxonomic subjects among the citation-based methods. Comparison of the taxonomies with roughly $10^5$ clusters shows that CC provides the least concentrated, and thus least accurate, solution. On a paper-by-paper basis, $DC^5$ has the highest Herfindahl index for 86.1% of the gold standard papers, and $BC^5$ is highest for 13.7%. $CC^5$ has the highest value for very few papers.

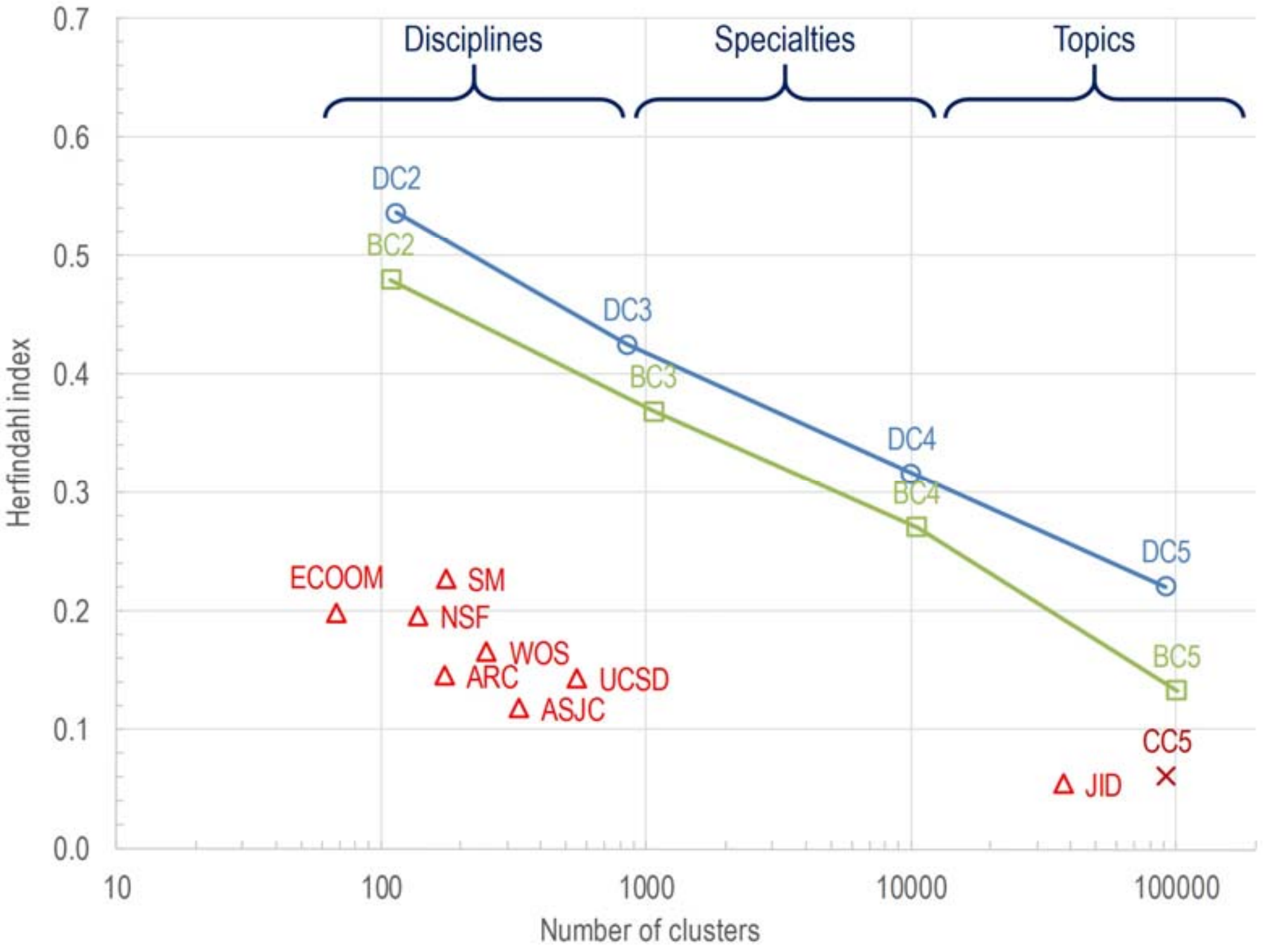

**Figure 2. Relationship between average Herfindahl value and number of clusters using different taxonomies for the references in 37,207 gold standard papers.**

There is also anecdotal evidence that CC is the least accurate method citation-based method from a historical perspective. We examined the titles of the top 5 clusters associated with the topic of citation analysis from the $DC^5$, $BC^5$, and $CC^5$ taxonomies, and found that of the three the $CC^5$ clusters were the least topically coherent; each cluster had different focal points in different cited time periods. The $DC^5$ and $BC^5$ clusters were both more topically focused. It is possible that the $CC^5$ clusters each contained multiple topic focuses because they were based on three citing years, while the $BC^5$ clusters were only based on a single citing year. However, the $DC^5$ clusters were the longest lived historically, and were still far more topically focused than the $CC^5$ clusters.

Figure 2 also shows that journal-based taxonomies are much less accurate than document-based taxonomies as representations of knowledge categories. This is particularly true for the case when each journal is considered as its own knowledge category – the JID method has a slightly lower concentration than even the $CC^5$ method. Excluding the JID method, the SM (Science-Metrix) journal schema has the highest accuracy while the ASJC journal schema has the lowest. Among the journal-based taxonomies, on a paper-by-paper basis, SM has the highest Herfindahl value for 43.2% of the papers, followed by NSF with 21.8% and ECOOM with 19.4%. Among the two vendor-provided classification systems, WOS scores better than ASJC (1.5% to 0.6%), which correlates well with a recent analysis of the two classification systems by Wang & Waltman (2015). However, neither vendor-provided classification system performs as well as some of those created by outside researchers. Figure 3 also suggests that a direct citation-based taxonomy of several hundred categories might be a better disciplinary representation than any of the journal cluster-based representations; $DC^3$ and $DC^2$ both have Herfindahl values well above those for all journal-based systems.

One thing that the representation in Figure 2 does not address is the possible effect of skewness in the cluster size distributions on the Herfindahl value. Cluster solutions that are more skewed have the potential to have higher Herfindahl values simply due to the presence of larger clusters at the top end of the distribution. Skewness values associated with cluster size distributions have been calculated for all taxonomies. The DC solutions are less skewed ($DC^{2-5}$: 1.42, 1.67, 1.65, 2.45) than the BC solutions ($BC^{2-5}$: 2.39, 4.20, 5.04, 3.89) at all levels, and thus do not have higher Herfindahl values due to skewness. In fact, the opposite may be true – the BC values may be slightly inflated due to skewness. Among journal-based taxonomies, the two highest scoring schema (SM and ECOOM) also have the two lowest skewness values. Thus, skewness does not appear to negatively affect our conclusions.

Overall, the data suggest that document-based taxonomies are more accurate than journal-based taxonomies in representing the structure of scientific and technical knowledge. In addition, among citation-based methods, direct citation creates the most accurate taxonomies at the document level. Thus, we feel very comfortable in claiming that direct citation creates the most accurate taxonomy of knowledge at the topic level.

***Detailed example***

Figure 2 provides clear evidence of the relative accuracies of each taxonomy. The accuracy and utility of a particular taxonomy, however, is best shown through an example. Science overlay maps were recently introduced by Rafols, Porter & Leydesdorff (2010), hereafter referred to as Rafols2010, in a paper that is one of the gold standards. It is located in cluster 9524 in the $DC^5$ taxonomy, contains 103 references of which 92 are available in $DC^5$. Rafols2010 is the gold standard paper that is most related to the current study, and will thus serve as our exemplar.

Table 7 shows the numbers of references from Rafols2010 that are found in the top four ranked clusters for each of the taxonomies. Large numbers of references are concentrated in the top-ranked cluster for the DC and BC taxonomies, with much lower numbers of references in the second ranked cluster. Despite similar numbers of references in top ranked clusters, Herfindahl values for BC taxonomies are lower than those for DC taxonomies because of lower coverage. In addition, one can see the hierarchical nature of the DC and BC solutions, with numbers of references in the top-ranked cluster increasing as one moves up the hierarchy (from level 5 to level 2). In contrast, the $CC^5$ taxonomy has no dominant cluster; the same small number of references occur in each of the top three ranked clusters, and $CC^5$ has a correspondingly low Herfindahl index. The journal-based taxonomies differ from the document-based taxonomies in that most of them have 10 or more references in at least two clusters. Thus, while they concentrate references more than does the $CC^5$ taxonomy, they are far less concentrated than the DC and BC taxonomies.

**Table 7. Distribution of references from Rafols2010 using different taxonomies.**

| | #Ref-1 | #Ref-2 | #Ref-3 | #Ref-4 | Herf |
|---|---|---|---|---|---|
| $DC^5$ | 44 | 8 | 4 | 2 | 0.244 |
| $DC^4$ | 61 | 7 | 2 | 2 | 0.459 |
| $DC^3$ | 68 | 3 | 3 | 2 | 0.551 |
| $DC^2$ | 71 | 3 | 3 | 2 | 0.600 |
| $CC^5$ | 7 | 7 | 7 | 4 | 0.030 |
| $BC^5$ | 39 | 2 | 1 | 1 | 0.186 |
| $BC^4$ | 54 | 2 | 1 | 1 | 0.349 |
| $BC^3$ | 58 | 3 | 2 | 2 | 0.403 |
| $BC^2$ | 59 | 7 | 2 | 2 | 0.421 |
| ASJC | 8.4 | 7.2 | 5.7 | 5.3 | 0.035 |
| UCSD | 25 | 14 | 9 | 2 | 0.112 |
| SM | 33 | 21 | 5 | 1 | 0.188 |
| ARC | 28 | 9 | 6 | 4 | 0.117 |
| ECOOM | 20 | 18 | 5.5 | 5.5 | 0.099 |
| WOS | 13.5 | 10 | 5 | 4.5 | 0.048 |
| NSF | 40 | 9 | 4 | 2 | 0.205 |
| JID | 20 | 11 | 9 | 6 | 0.082 |

The following description will focus on the distribution of references in the $DC^5$ taxonomy. This is the taxonomy that we favor using because it is created by the most accurate method, and because it addresses the topic level (~$10^5$ clusters), rather than discipline level. Our experience has been that the majority of science policy questions of interest to us (e.g., which topics are emerging, declining, receiving funding, etc.) are best answered at the topic level. However, we are also aware that others feel that there are too many small clusters at the topic level, and as such, favor the specialty level (Ruiz-Castillo & Waltman, 2015). The proper level of granularity likely depends on the specific question being asked, and is a question that we do not address in this study.

The two clusters concentrating the largest numbers of references from Rafols2010 are $DC^5$-9524 and $DC^5$-432, with 44 and 8 references, respectively. Both of these clusters are characterized in Figure 3 using word clouds (generated from all titles in the cluster, not just the titles referenced by Rafols2010) and growth curves. $DC^5$-9524 focuses on science mapping as part of the process of analyzing research activities, and reports on advances in citation, co-occurrence and network analysis related to community detection and the structure of science, along with related advances in visualization. Given its primary focus, it is no surprise that Rafols2010 appears in this cluster, and that a large fraction of its references (44/92) are found here as well. While Rafols2010 uses subject categories, rather than documents or journals, as the basis for mapping, the underlying co-occurrence analysis technique employed, the claim that the map represents a typology of knowledge, and the visualization technique used to representing the resulting network are all historical artifacts belonging to this cluster. The advances made in Rafols2010 (the subject category map and the overlay technique) build directly on these historical features.

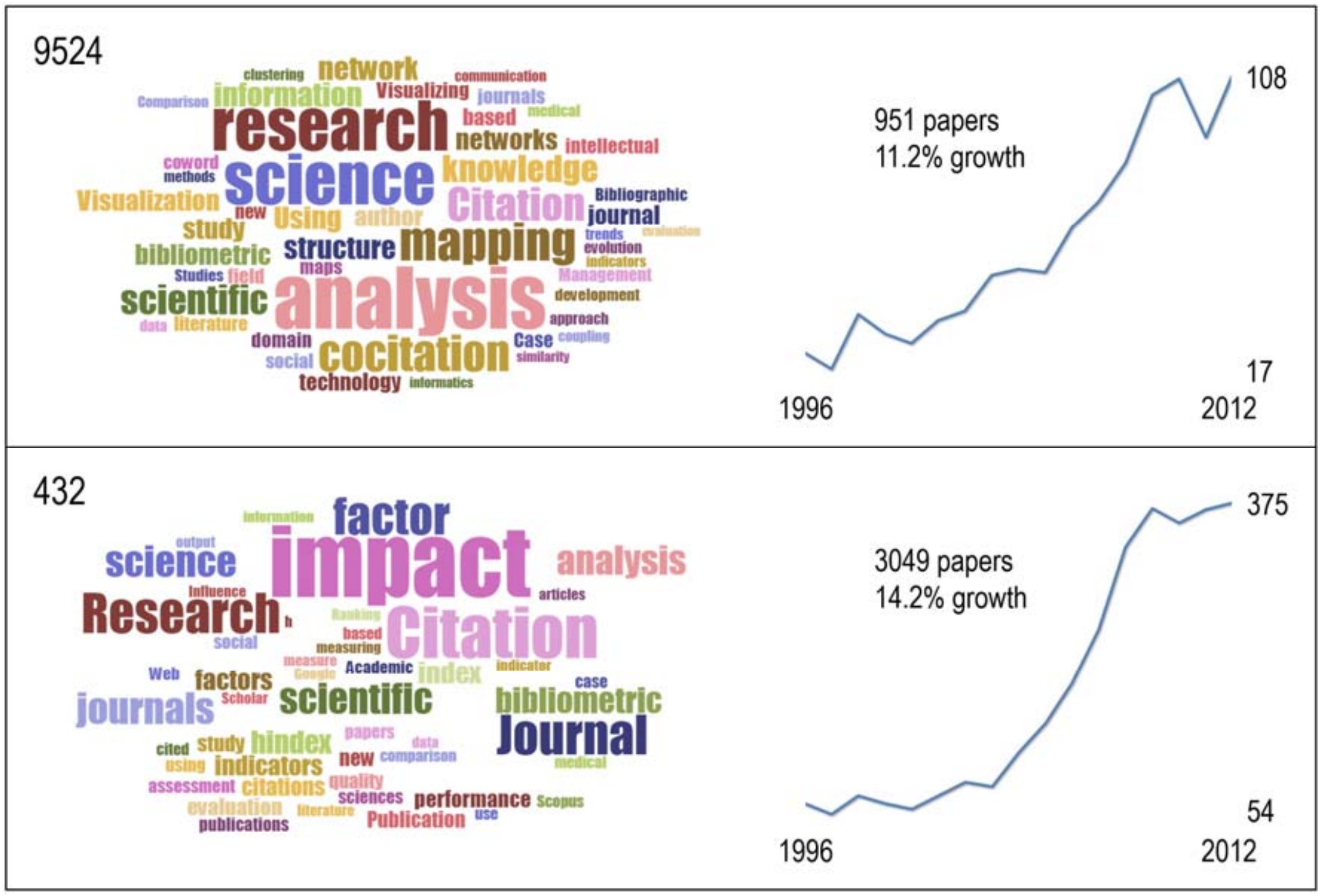

**Figure 3. Word clouds and publication growth curves for clusters DC[5]-9524 and DC[5]-432. Sizes and annual growth rates are included.**

DC[5]-9524 is currently producing about 100 papers annually, of which about 75 are articles or reviews, and 25 are conference papers. An average cluster of this size would include 2-3 papers per year with at least 100 references. However, Rafols2010 is the only paper in DC[5]-9524 with over 100 references, not only for 2010, but for the entire time period from 1996-2012. It is not clear why there are so few papers with large numbers of references in this cluster, but it may have something to do with the fact that clusters with advances in basic science are more likely to feature large numbers of review papers than clusters focused on methodologies.

DC[5]-432 is the cluster cited second most by Rafols2010; it focuses on impact and metrics, and is more than three times as large as DC[5]-9524. This cluster is based on the work of Garfield and Price in the sense that it builds on their early articles arguing that citations are evidence of impact. It includes literature about journal impact factors, research evaluation studies of institutions and nations, and more recently, the h-index and associated citation-based metrics. Two articles from 2010 with 100 references appear in this cluster – a review on the h-index by Egghe (2010) and a comparison of h-indexes by García-Pérez (2010). A total of 20 articles with at least 100 references appear in this cluster over the time period from 1996-2012. It is also interesting that 91 documents from *Science* and *Nature* appear in $DC^5$-432, while $DC^5$-9524 contains only three. This correlates with the controversial nature (in the popular sense) of the topics of metrics and impact, and the relative lack of controversy associated with the academic development of science mapping (and related) methodologies.

## Summary and Implications

Research planning and evaluation are key activities in current society. Practitioners are found across government, industry, academia, and the non-profit world, and decisions are made at agency, administrative, and researcher levels alike. Such decisions require an accurate picture of the structure (taxonomy) of science and technology. This accurate picture has been sought since the days of Garfield and Price, and only now, after 50 years, do we have the data and computing resources to bring the full taxonomy of knowledge into focus. The purpose of this paper has been to take one more step toward being able to accurately depict the taxonomy of scientific and technical knowledge from a socio-cognitive perspective.

Using cited references from papers with at least 100 references, which we propose as gold standards for taxonomic subjects, we have compared the accuracies of topic-level taxonomies based on the clustering of documents using direct citation, bibliographic coupling, and co-citation. Several discipline-level taxonomies based on journal clusters have also been compared. The data show that direct citation concentrates references at a higher level than either bibliographic coupling or co-citation. Using the assumption that higher concentrations of references denote more accurate clusters, direct citation thus provides a more accurate representation of the taxonomy of scientific and technical knowledge than either bibliographic coupling or co-citation. This is not to say that every paper is perfectly assigned using the direct citation approach – to assume such would be naïve – but rather that on the whole, the clusters in the direct citation taxonomies are better than the clusters in the other taxonomies.

One argument that might be made against this result is that direct citation might have turned out to be the most accurate method simply because the measure of accuracy is based on direct citations from gold standard articles. Admittedly, the measure of accuracy is not independent of any of the citation-based processes. Thus, such circularity of reasoning is to some extent unavoidable unless one can identify a truly independent measure of relatedness between articles. Nevertheless, it is also true that direct citations are the most reasonable way to quantify accuracy simply because they are first-order (direct) indicators of the relatedness of publications, while co-citation and bibliographic coupling links are second-order (and thus indirect) indicators of relatedness. Moreover, it is also true that the 37,207 gold standard papers had relatively little impact on the direct citation clustering – these papers were only involved in 5.3 million (0.9%) out of the 582 million links used in the direct citation calculation.

Direct citation, bibliographic coupling, and co-citation naturally produce solutions with different age distributions. As shown in Figure 4, a direct citation taxonomy that includes cited non-source documents contains far more of the historical literature (on a fractional basis) than the other taxonomies. In contrast, bibliographic coupling is strongly weighted to the most recent references. This may explain why bibliographic coupling was found to be far more accurate than direct citation in our previous study (Boyack & Klavans, 2010), which used a recent five-year window of data, while direct citation is most accurate in this study, which focuses on long-term historical clusters.

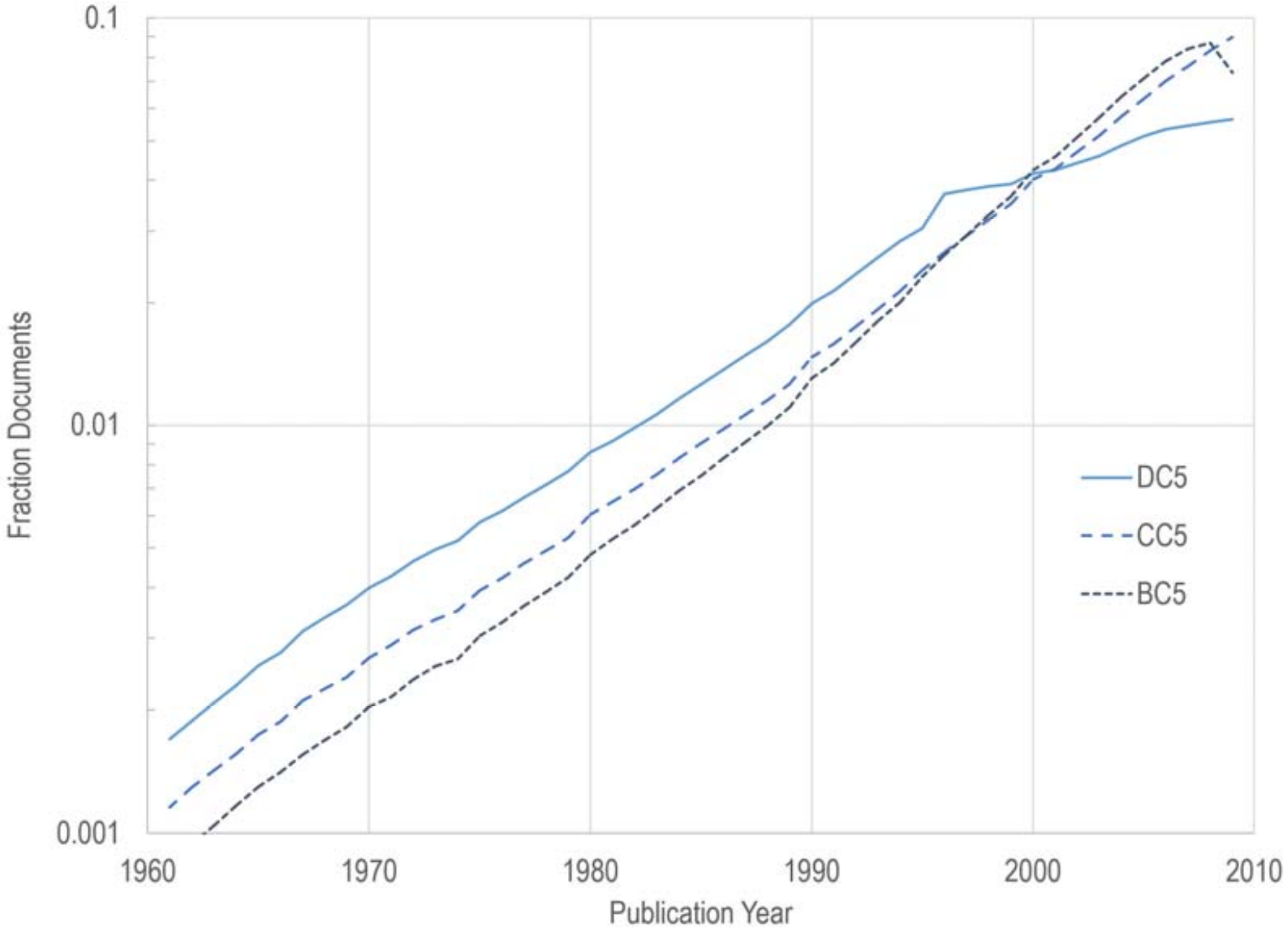

**Figure 4. Comparison of DC, CC, and BC document distributions by year.**

Regarding journal classification systems, this study compared existing systems with document-based structures with similar numbers of clusters, and found that document-based taxonomies provide a more accurate representation of disciplines than do journal-based taxonomies. A similar conclusion was reached by Ruiz-Castillo & Waltman (2015), and the Leiden Ranking has accordingly been changed – it now uses an algorithmically constructed taxonomy rather than journal subject categories as its taxonomic basis. Furthermore, the Leiden taxonomy was constructed using the same direct citation methodology shown here to be the most accurate.

These findings have multiple implications for the research streams in $DC^5$-9524 and $DC^5$-432, which are, as in Rafols2010, the two clusters referenced most by our work. Of the 49 references in our reference list, 15 are too recent to appear in the $DC^5$ model, and of the remaining 34 references, 19 and 5 are to clusters $DC^5$-9524 and $DC^5$-432, respectively.

$DC^5$-9524 focuses on science mapping, yet the current paper, which clearly belongs to the tradition associated with $DC^5$-9524, does not describe any science mapping methodology in detail (those details are available in previous papers), and it does not show a map. Rather, its purpose is solely to compare the accuracies of existing mapping methodologies – to 'get the nodes right.' Our findings, which may be difficult for many to accept due to tradition, suggest that use of journals or journal clusters as a way to identify areas of knowledge is a poor choice, and call into question the utility of studies that use journals or journal categories as a basis for research evaluation, diffusion, innovation, and a host of other applications. For example, these results lead us to question the assumption by Uzzi et al. (2013) that journals are suitable proxies for knowledge areas and that journal-journal relationships can thus represent atypical relationships. We replicated Uzzi's study (Boyack & Klavans, 2014a) and found that the dominant signals of atypical relationships were due to highly influential journals that are not representative of homogeneous knowledge areas (e.g., *Science*, *Nature*, *PNAS*, etc.). Does a combination of atypical and typical relationships signal innovative activity? We believe so, but also suggest that topics, instead of journals, would have been a far better basis for testing this hypothesis.

As another example, we take a closer look at Rafols2010. Is this a multidisciplinary study or not? Using several of the journal-based taxonomies and distribution of references from Table 7, one might conclude that the paper is multidisciplinary since the top two disciplines contain nearly equal numbers of references (e.g., UCSD - 25:14, SM - 33:21 or ECOOM - 20:18). However, having read the paper, we would conclude that the paper is not multidisciplinary (in much the same way we believe that our paper is not multidisciplinary) because it builds primarily on a set of traditions from a single topic. Direct citation analysis may, in fact, be detecting disciplines as they presently exist because they build directly on communication and influence patterns in the literature.

The findings of this study have also caused us to re-evaluate our own work. For many years we have been strong proponents of co-citation analysis – the majority of our published work over the past decade has been related in one way or another to use of co-citation, and earlier efforts to measure accuracy suggested that co-citation analysis was competitive (Boyack & Klavans, 2010). However, it is clear from this study that co-citation is inferior to direct citation if the goal is to create a knowledge taxonomy. Not only is the coverage lower, but the accuracy is much lower. While it is possible that co-citation might be improved using different windows and thresholds, we no longer intend to pursue co-citation as an avenue of research. In its stead, we now believe that research efforts should focus on Price's distinction between taxonomic subjects, which are stable and incorporate the historical record, and research fronts, which can change rapidly, and, in a sense, obliterate history as time progresses. Improvements in both should be easier to detect by using the gold standards suggested in this paper. Analyses into the innovativeness of papers, researchers, institutions and nations should be more accurate if one shifts to a more accurate method for detecting topics.

The researchers associated with $DC^5$-432 are very concerned that research evaluations, using citation data, are creating distortions in the research allocation process. These concerns cover problems with evaluating the strength and innovativeness of individual papers, researchers, departments, institutions and nations. Of particular concern is the current trend towards rewarding past performance – funding researchers and institutions with the highest impact – and a failure to encourage more innovative research going forward.

We share this concern, and believe that citationists should be adopting the most accurate and comprehensive metrics possible. Stated rhetorically, if government agencies are currently spending millions of dollars on evaluation and planning for the purpose of making more informed decisions about the allocation of billions of research dollars, is it reasonable to use metrics that are proven to be inferior simply because they are based on tradition and convenience? The answer to this is clearly no. Rather, use of the most accurate methods for characterizing knowledge taxonomies and research fronts is critical to reducing the distortions in the research allocation process. This paper shows that direct citation produces a very accurate taxonomy of all of science from socio-cognitive and historical perspectives. We suggest that this taxonomy would be a suitable basis for decision making, and that metrics of innovation based on this taxonomy would be an improvement over traditional metrics. For example, we recently developed a new topic-level innovation metric using the $DC^5$ direct citation taxonomy, and found a very strong correlation between this metric and STAR METRICS® funding data from U.S. funding agencies (Boyack & Klavans, 2015). Additional study is needed. Nevertheless, these results reflect the potential utility of such models in informing decisions about resource allocation and research strategy. Further, we envision that an accurate taxonomy could be used to identify those topics where 'jumps' and 'bridges' (Foster, Rzhetsky, & Evans, 2015) are most likely to occur in the future, and that this might help us identify the actors (Latour, 1987) – scientists, funders, chemicals, equipment, etc. – that can be most fruitfully leveraged for innovation. We invite others to join us in this research effort.

## Acknowledgments

We thank Henry Small and Ludo Waltman for instructive comments on this manuscript. Ludo, in particular, pointed out the possibility of an error in our original ASJC calculations (which was indeed an error), for which we are very grateful. We also appreciate the very helpful comments from two additional reviewers.

## References

Aksnes, D. W. (2003). Characteristics of highly cited papers. *Research Evaluation, 12*(3), 159-170.

Archambault, E., Caruso, J., & Beauchesne, O. (2011). Towards a multilingual, comprehensive and open scientific journal ontology. *Proceedings of the 13th International Conference of the International Society for Scientometrics and Informetrics*, 66-77.

Börner, K., Klavans, R., Patek, M., Zoss, A. M., Biberstine, J. R., Light, R. P., et al. (2012). Design and update of a classification system: The UCSD map of science. *PLoS ONE, 7*(7), e39464.

Boyack, K. W. (2009). Using detailed maps of science to identify potential collaborations. *Scientometrics*, *79*(1), 27-44.

Boyack, K. W., & Klavans, R. (2010). Co-citation analysis, bibliographic coupling, and direct citation: Which citation approach represents the research front most accurately? *Journal of the American Society for Information Science and Technology, 61*(12), 2389-2404.

Boyack, K. W., & Klavans, R. (2014a). Atypical combinations are confounded by disciplinary effects. *19th International Conference on Science and Technology Indicators*.

Boyack, K. W., & Klavans, R. (2014b). Creation of a highly detailed, dynamic, global model and map of science. *Journal of the Association for Information Science and Technology, 65*(4), 670-685.

Boyack, K. W., & Klavans, R. (2014c). Including non-source items in a large-scale map of science: What difference does it make? *Journal of Informetrics, 8*, 569-580.

Boyack, K. W., & Klavans, R. (2015). Is the most innovative research being funded? *20th International Conference on Science and Technology Indicators*.

Boyack, K. W., Klavans, R., Small, H., & Ungar, L. (2014). Characterizing the emergence of two nanotechnology topics using a contemporaneous global micro-model of science. *Journal of Engineering and Technology Management, 32*, 147-159.

Boyack, K. W., Newman, D., Duhon, R. J., Klavans, R., Patek, M., Biberstine, J. R., et al. (2011). Clustering more than two million biomedical publications: Comparing the accuracies of nine text-based similarity approaches. *PLoS One, 6*(3), e18029.

Davidson, G. S., Wylie, B. N., & Boyack, K. W. (2001). Cluster stability and the use of noise in interpretation of clustering. *Proceedings IEEE Information Visualization 2001*, 23-30.

Egghe, L. (2010). The Hirsch index and related impact measures. *Annual Review of Information Science and Technology, 44*, 65-114.
Foster, J. G., Rzhetsky, A., & Evans, J. A. (2015). Tradition and innovation in scientists' research strategies. *American Sociological Review, 80*(5), 875-908.
Franklin, J. J., & Johnston, R. (1988). Co-citation bibliometric modeling as a tool for S&T policy and R&D management: Issues, applications, and developments. In A. F. J. van Raan (Ed.), *Handbook of Quantitative Studies of Science and Technology* (pp. 325-389). North-Holland: Elsevier Science Publishers, B.V.
García-Pérez, M. A. (2010). Accuracy and completeness of publication and citation records in the Web of Science, PsycINFO, and Google scholar: A case study for the computation of h indices in psychology. *Journal of the American Society for Information Science and Technology, 61*(10), 2070-2085.
Garfield, E. (1955). Citation indexes for science: A new dimension in documentation through association of ideas. *Science, 122*, 108-111.
Garfield, E., Sher, I. H., & Torpie, R. J. (1964). *The Use of Citation Data in Writing the History of Science*. Philadelphia: Institute for Scientific Information.
Glänzel, W., & Schubert, A. (2003). A new classification scheme of science fields and subfields designed for scientometric evaluation purposes. *Scientometrics, 56*(3), 357-367.
Hric, D., Darst, R. K., & Fortunato, S. (2014). Community detection in networks: Structural communities versus ground truth. *Physical Review E, 90*, 062805.
Ioannidis, J. P. A., Boyack, K. W., & Klavans, R. (2014). Estimates of the continuously publishing core in the scientific workforce. *PLOS One, 9*(7), e101698.
Ioannidis, J. P. A., Boyack, K. W., Small, H., Sorensen, A. A., & Klavans, R. (2014). Is your most cited work your best? *Nature, 514*, 561-562.
Kessler, M. M. (1963). Bibliographic coupling between scientific papers. *American Documentation, 14*(1), 10-25.
Kessler, M. M. (1965). Comparison of the results of bibliographic coupling and analytic subject indexing. *American Documentation, 16*(3), 223-233.
Klavans, R., & Boyack, K. W. (2006). Quantitative evaluation of large maps of science. *Scientometrics, 68*(3), 475-499.
Klavans, R., & Boyack, K. W. (2010). Toward an objective, reliable and accurate method for measuring research leadership. *Scientometrics, 82*(3), 539-553.
Kuhn, T. S. (1962). *The Structure of Scientific Revolutions* (1st ed.). Chicago: University of Chicago Press.
Kuhn, T. S. (1970). *The Structure of Scientific Revolutions* (2nd ed.). Chicago: University of Chicago Press.
Latour, B. (1987). *Science in Action*. Cambridge: Harvard University Press.
Leydesdorff, L. (1987). Various methods for the mapping of science. *Scientometrics, 11*(5-6), 295-324.
Leydesdorff, L., Rafols, I., & Chen, C. (2013). Interactive overlays of journals and the measurement of interdisciplinarity on the basis of aggregated journal-journal citations. *Journal of the American Society for Information Science and Technology, 64*(12), 2573-2586.
Martin, S., Brown, W. M., Klavans, R., & Boyack, K. W. (2011). OpenOrd: An open-source toolbox for large graph layout. *Proceedings of SPIE - The International Society for Optical Engineering, 7868*, 786806.

Newman, M. E. J. (2006). Modularity and community structure in networks. *Proceedings of the National Academy of Sciences of the USA, 103*(23), 8577-8582.
Newman, M. E. J., & Girvan, M. (2004). Finding and evaluating community structure in networks. *Physical Review E, 69*, 026113.
Pautasso, M. (2013). Ten simple rules for writing a literature review. *PLoS Computational Biology, 9*(7), e1003149.
Price, D. J. D. (1965). Networks of scientific papers. *Science, 149*, 510-515.
Price, D. J. D., & Gürsey, S. (1975). Studies in scientometrics I: Transience and continuance in scientific authorship. *Ci. Inf. Rio de Janeiro, 4*(1), 27-40.
Rafols, I., Porter, A. L., & Leydesdorff, L. (2010). Science overlay maps: A new tool for research policy and library management. *Journal of the American Society for Information Science and Technology, 61*(9), 1871-1887.
Ruiz-Castillo, J., & Waltman, L. (2015). Field-normalized citation impact indicators using algorithmically constructed classification systems of science. *Journal of Informetrics, 9*, 102-117.
Small, H. (1973). Co-citation in the scientific literature: A new measure of the relationship between two documents. *Journal of the American Society for Information Science, 24*, 265-269.
Small, H. (1999). Visualizing science by citation mapping. *Journal of the American Society for Information Science, 50*(9), 799-813.
Small, H., Boyack, K. W., & Klavans, R. (2014). Identifying emerging topics in science and technology. *Research Policy, 43*, 1450-1467.
Small, H., & Griffith, B. C. (1974). The structure of scientific literatures, I: Identifying and graphing specialties. *Social Studies of Science, 4*, 17-40.
Small, H., Sweeney, E., & Greenlee, E. (1985). Clustering the Science Citation Index using co-citations. II. Mapping science. *Scientometrics, 8*(5-6), 321-340.
Uzzi, B., Mukherjee, S., Stringer, M., & Jones, B. (2013). Atypical combinations and scientific impact. *Science, 342*, 468-472.
Waltman, L., & van Eck, N. J. (2012). A new methodology for constructing a publication-level classification system of science. *Journal of the American Society for Information Science and Technology, 63*(12), 2378-2392.
Waltman, L., & van Eck, N. J. (2013). A smart local moving algorithm for large-scale modularity-based community detection. *European Physical Journal B, 86*, 471.
Wang, Q., & Waltman, L. (2015). Large-scale analysis of the accuracy of the journal classification systems of Web of Science and Scopus. *arXiv:1511.00735 [cs.DL]*.
Woodward, A. M. (1977). The roles of reviews in information transfer. *Journal of the American Society for Information Science, 28*, 175-180.

---

[i] Selye, H. (1946). The general adaptation syndrome and the diseases of adaptation. *Journal of Clinical Endocrinology*, 6, 117-231.

[ii] Asimov, I. (1964). *The genetic code. The story of DNA - The chain of life.* London: John Murray.

[iii] Unpublished work.

[iv] https://www.elsevier.com/__data/assets/excel_doc/0015/91122/title_list.xlsx

[v] http://sci.cns.iu.edu/ucsdmap/data/UCSDmapDataTables.xlsx

[vi] http://science-metrix.com/files/science-metrix/sm_journal_classification_105_1_0.xls

[vii] http://tqft.net/math/ERA2015.csv

[viii] https://www.kuleuven.be/research/bibliometrics/ecoomjournallist.xlsx

[ix] June 2015 WoS journal file obtained from Ludo Waltman, CWTS; active journals only

[x] File obtained from Lawrence Burton (now retired) at NSF in 2007